# Cloud Storage Forensic: hubiC as a Case-Study

Ben Blakeley, Chris Cooney, Ali Dehghantanha, Rob Aspin
School of Computing, Science, and Engineering, University of Salford
Manchester, United Kingdom
ben@benblakeley.com, christopherpiercecooney@gmail.com, A.Dehghantanha@Salford.ac.uk, R.Aspin@salford.ac.uk

*Abstract*— In today's society where we live in a world of constant connectivity, many people are now looking to cloud services in order to store their files so they can have access to them wherever they are. By using cloud services, users can access files anywhere with an internet connection. However, while cloud storage is convenient, it also presents security risks. From a forensics perspective, the increasing popularity of cloud storage platforms, makes investigation into such exploits much more difficult, especially since many platforms such as mobile devices as well as computers are able to use these services. This paper presents investigation of hubiC as one of popular cloud platforms running on Microsoft Windows 8.1. Remaining artefacts pertaining different usage of hubiC namely upload, download, installation and uninstallation on Microsoft Windows 8.1are presented.

## I. INTRODUCTION

Cloud Computing may be defined as *"a model for enabling ubiquitous, convenient, on-demand network access to a shared pool of configurable computing resources (e.g., networks, servers, storage, applications, and services) that can be rapidly provisioned and released with minimal management effort or service provider interaction."* [1] and has provided a powerful platform for criminal use [2]. It has been used to store, share and transfer malicious content. Ergo, investigations of cloud platforms rise as a priority for forensic experts. Investigation of cloud storage services poses new challengesarising from the federated nature of information which presents multiple legal issues [3].

Utilizing a forensics investigation to analyze Amazon S3, Dropbox, Evernote and Google Docs on Motorola Droid with Android v2.2.2, iPhone 4 with iOS v4.3.5, Mac and Windows Machines [4] identified a variety of digital remnants including; username, downloaded and uploaded filenames. Moreover, investigations of Dropbox, Google Docs, PicasaWeb and Flicker,carried out on Windows 7 PCs, have recovered artefacts of web browser and desktop applications [5]. Dykstra and Sherman analysed export features of Amazon EC2 to illustrate a cloud evidence collection method using different forensics tools such as EnCase and AccessData FTK [6]. Hale also recovered a variety of artifacts, such as the installation path and registry modifications [7], from an Amazon cloud drive on Windows XP when analyzing both server and client sides of a private cloud storage service [9], authentications credentials, file contents and timestamps were found, allowing for the full recovery of server files. Oother investigations, highlighted that, despite the challenge of obtaining logs from cloud services, these services held substantial worth to a forensic investigator [11] and a tool was provided for accessing the logs on a cloud service.

Martini and Choo proposed a collection process, consisting of six steps, for investigating VMware vCloud and showed that by using vCloud's REST API, a variety of evidential data could be collected [14]. Similarly, a cloud data imager was developed to collect evidences from Google Drive, Dropbox and Skydrive [15], providing interaction (browsing and imaging) with folder trees.

Google Drive was employed as a case study to test and demonstrate a forensic analysis framework [16]. This was subsequently tested against SkyDrive [17], Dropbox [18], SugarSync [19] and UbuntuOne [13] cloud services.

A summary of cloud storage forensic research is outlined in Table I.

TABLE I. A SNAPSHOT OF EXISTING CLOUD FORENSICS RESEARCH

| General Cloud Research | Public cloud | Private cloud |
|---|---|---|
| Mell & Grance [1] | Amazon Cloud Drive [7] | Dropbox, Box and Sugarsync [8] |
| Daryabar et al. [2] | XtreemFS [12] | Dropbox [18] |
| Damshenas et al [3] | Ubuntu One [13] | Sugarsync [19] |
| Chung et al. [4] | Dropbox, Google Drive and Microsoft SkyDrive [15] | |
| Marturana et al. [5] | Google Drive [16] | |
| Dykstra & Sherman [6] | Skydrive [17] | |
| Martini & Choo [9] | | |
| Wen et al. [10] | | |
| Zawoad et al [11] | | |
| Martini & Choo [14] | | |
| Dezfoli et al. [20] | | |
| Daryabar et al. [21] | | |
| Damshenas [22] | | |

hubiC is a rapidly growing cloud storage service, created by OVH (On Vous Héberge) at the end of 2011. This cloud service celebrated 500,000 unique users on 27th March 2015, with ambitions to achieve 1,000,000 users by 2020. hubiC was originally solely available to users in France until 2013, after which the ability to register accounts for international users was offered. However, literature suggests that investigation of hubiC as a cloud storage platform has been left untouched. Therefore, the research reported here aims to provide insight into the remnants of data left behind by the hubiC cloud service on Windows 8.1 (the current active, stable mass

consumer operating system). This research paper aims to answer following questions:
- What data can be recovered on the hard drive of a Windows 8.1 machine after the use of the hubiC cloud storage service?
- What data can be recovered from the physical memory (RAM) of a Windows 8.1 machine after the use of the hubiC cloud storage service?
- What data is transmitted in the network traffic during communication with the hubiC cloud storage service during the upload and download of files?

The following sections present the research experiment setup, followed by detailed discussion of detected remnants. The paper concludes providing a summativereview of findings and proposals of future works.

## II. EXPERIMENT SETUP

The experiment was conducted over various VMs. A base VM was created, containing a clean installation of Windows 8.1.9.3.9600. This VM was then cloned to the following VMs:
**The Access VM (1.1)** – To detect remnants of accessing and logging into hubiC via the web browser.
**The Upload VM (1.2)** – To investigate evidences of uploading files to hubiC.
**Download/Open VM (1.3)** – To detect evidences during the downloading and opening of a file from the web browser.
**Delete VM (1.4)** – Investigating file remnants left behind when a file is deleted via the web browser.
**Install Desktop VM (1.5)** – Searching for evidences after installation of the hubiC desktop application.
**Upload Desktop VM (1.5.1)** – Searching for file remnants after upload of a test file via the hubiC desktop application.
**Uninstall Desktop VM (1.5.1.1)** – Searching for evidences of hubiC desktop application existence and test files.
**Download Desktop VM (1.5.2)** – Investigating evidences available after using hubiC desktop application to download a test file.
**Delete Desktop VM (1.5.2.1)** – Searching for file remnants after deleting test file using hubic desktop application.
Internet Explorer v11.0.9600.16384, Google Chrome v40.0.2214.115 and Mozilla Firefox v35.0.1 were used during web-browser based experiments. When using Chrome, ChromeCacheView V1.66 was used to view data remnants left behind when accessing the hubiC website. Additionally, the HxD hex editor V1.7.7.0 was used for the memory forensics portion of the investigation. Email 22 from the */maildir/delainey-d/inbox* directory of the Enron email dataset(http://www.cs.cmu.edu/~enron/)( 10/04/2015), was used to upload to the hubiC server. The network traffic was investigated using Wireshark 1.12.4.

## III. RESULTS AND ANALYSIS

In this section, results of investigating the Access VM (1.1), Upload VM (1.2), Download/Open VM (1.3), Delete VM (1.4), Install Desktop VM (1.5), Upload Desktop VM (1.5.1), Uninstall Desktop VM (1.5.1.1), Download Desktop VM (1.5.2) and Delete Desktop VM (1.5.2.1), including memory forensics and analysis of temporary and log files, which provide insight into the internal operations of hubiC are presented. Network monitoring was also conducted during these investigations.

### A. Network Traffic Analysis

During analysis of the network traffic, it was observed that when the initial request was made to www.hubic.com, a redirect to https on port 8080 was returned. This indicates that there is no plaintext traffic accepted by the hubiC website, a subsequent redirection response is then sent, with the status code 302 (moved temporarily), which finally redirects to https://www.hubic.com/en.

All communication between the hubiC client and server applications was encrypted using TLSV1, preventing successful extraction of data of any sniffed traffic. Therefore, no clear indication of the e-mail address or the password could be detected within the login stream. Once login was completed, a session id was created, although this was not saved into the cookies. This value was a combination of "hubiC.com" and a timestamp. All involved domain names and their respective IP address were analyzed. Most of the IP addresses did not have an available DNS mapping. During analysis, we managed to infer the certificate type, by following the TCP packets in transit using Wireshark as shown in Figure 1.

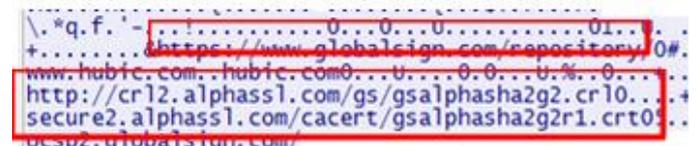

Figure 1 – Certificate Data

During uninstallation, packets were found to be exchanged with the hubiC server. The TCP conversation revealed that there was no communication between the hubiC client and server during the uninstallation process, indicating that all changes were performed locally.

During the upload of the file, hubiC communicated with 37.59.76.98, which, again, is an OVH server without a domain. Initiation of upload traffic streams are tagged with "hubiC.ovh.net" as shown in Figure 2. Investigation of uploaded traffic shows utilization of compression algorithms before files are sent. For example, corresponding stream of a 1318 bytes uploaded file was just 1307 bytes.

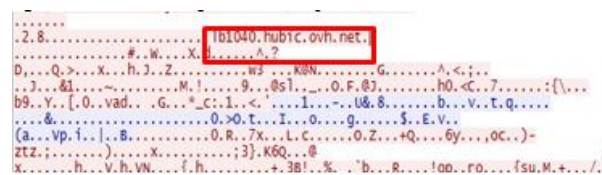

Figure 2 – hubiC string in TCP stream.

During the upload, a GIF file was found that was named "prepareUpload.gif". This provided information about what was happening during the usage of the machine, but did not present any information about the file itself.

## B. Analysis of Access VM (1.1)

During analysis of the Access VM (1.1), the browser cache was used to attempt to obtain knowledge of when the access occurred, along with temporary internet files (such as browsing history and cookies). Using the Chrome cache viewer, Javascript, HTML and Image files were obtained that were downloaded during transactions with the hubiC browser interface (Figure 3). Analysis also revealed the e-mail address and password of the user that was used to log in to hubiC (Figure 4). This was found within the process memory of Internet Explorer using "password" as a keyword, by searching directly through the memory using HxD hex editor. Additionally, a remnant of the page title was found in memory (Figure 5).

## C. Analysis of Upload VM (1.2)

For VM 1.2 the remnants that were left behind during the uploading of a file. via the web browser, to the hubiC server, were analysed.

SHA1 checksums of the file before and after the upload were exactly the same.

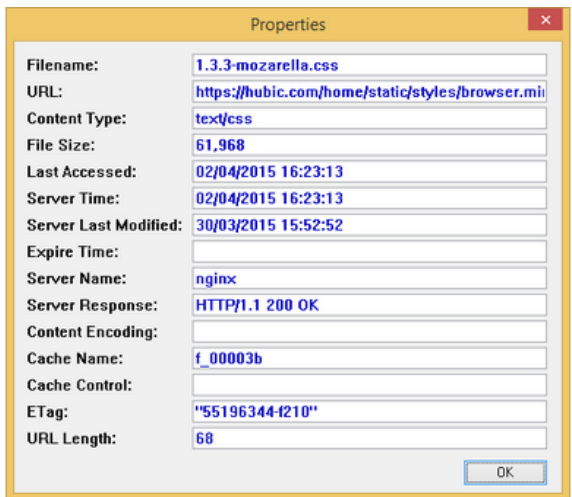

Figure 3 – Example of content in CacheMemory

Figure 4 - E-Mail address and password of the user in memory

Figure 5 – Page title found in memory

Searching through the chrome cache files in *%USERPROFILE%/AppData/Roaming/Google/Chrome* led to detection of an additional "prepareUpload.gif" file. Opened this file in a hex editor revealed that the image was a single pixel only. Examining the bytes of this GIF file did not reveal any relevant information hidden inside the image, or inside the image slack space. When the website was accessed from Google Chrome it is clear, from the system memory, that files were deposited in the chrome local storage folder with clear indication that the data resulted from accessing hubic.com.

In the metadata of the file, the e-mail file had the original created date. After the upload, the file was then analyzed to check if these values have been changed. The same file displayed no change to the metadata of the file, including last accessed timestamp.

Memory analysis of the upload VM revealed the uploaded file was available in the process space of Internet Explorer.

## D. Analysis of Download/Open VM (1.3)

The first assessment performed was the hash comparison. The hash comparison matches indicated that the contents of the file had not been changed during the download process. The file metadata was then compared with the original. As expected, , both the date created and date accessed were the time that the file was downloaded to the disk, however the date modified field was set to two seconds after.

During the memory analysis, the downloaded file was found in multiple locations. The first location represented the file that was loaded into memory by Internet Explorer during the download (the same file was not found on all other browsers),. the second copy located on the hard disk. These file locations were as expected.

## E. Analysis of Delete VM (1.4)

The bulk of the analysis came from investigation into the cache. While most browsers did not cache any files pertinent to communication with the hubiC web client, beyond HTML fragments, the Chrome Cache contained references to a gif file. In this instance, the file was named "delete". This file was uploaded when the delete command was sent to the server.

## F. Analysis of Install Desktop VM (1.5)

Using regedit.exe, the registry of the Install VM was scanned to find references to hubiC post installation. An assessment into the registry found entries at HKEY_CURRENT_USER/Software/OVH/hubiC-sync. After install, logs were present in AppData/Local/Temp. The names of these log files contained the time when installation was performed. The only configuration file of hubiC (hubiC.exe.config) was located at %programfiles%\OVH\hubiC. This file indicated some basic setup for how the application should parse the URL.

## G. Analysis of Upload Desktop VM (1.5.1)

The first stage was to obtain the process ID of the hubiC process. This was discovered using Resource Monitor. Once the application was started up, the log files were assessed to find relevant changes. The logs written to during installation were not changed, but new logs were created in the AppData/Roaming/hubiC folder. This folder contained the application.txt file, the log file, and hubiC.db file. hubiC.db was an SQLite file. This file was subsequently opened using the Firefox plugin SQLite Manager 0.8.3. An assessment of this database before an upload had occurred showed a collection of empty tables. These tables were named backupDirs, backupFileVersions, backupFiles, files, localHashCache, remoteHashCache and synchronizedDirs. This file was then uploaded and the state of the SQLite DB was monitored. After logging in to the hubiC client application, a new database was made named "hubiC.db-journal". Initially, hubiC.db was analysed. Data was now found in the files. Data was also found in the synchronizedDirs table as shown in Figure 6.

Figure 6 – New entry in table

Once a hubiC directory was chosen, three separate locations were placed into the files table. These folders corresponded to folders on the hubiC cloud server, indicating synchronization. When attempting to access the *hubiC.journal-db*, an error message was presented, indicating that the database file was either encrypted or corrupted. To combat this, the file was carved using a hex editor, however the encryption remained. As a way to glean information from this, the bytes were analyzed in an attempt to calculate the structure of the database table and identify any data contained. DDL syntax was found, indicating the structure of the database. This DDL syntax seemed to create the files table, in the same structure as found in the hubiC.db file.

An analysis of the file metadata revealed that there was no change to the original file during the upload process. Registry files were also observed during the upload process. The login details used are clearly stored at *HKEY_CURRENT_USER/Software/OVH/hubiC-sync*. The key "AccountEmail" contains the reference to the email address of the logged in user, along with SecureAuthToken, which is an oAuth 2.0 secure token, granting the hubiC software access to the API.

Memory analysis was again carried out on the virtual machine using a hex editor. Within this, the e-mail address used to log into the application was discovered as shown in Figure 7. In addition to this, the name of the file uploaded to the hubiC server was found in memory.

Figure 7 – Email account found in memory

## H. Analysis of Uninstall Desktop VM (1.5.1.1)

The uninstallation was performed using the Windows uninstaller: hubiC does not have its own uninstaller application.

Analysis of the log files stored in *AppData/Local/Temp* revealed they were not removed during the uninstallation process; however they contained no information within them about the application being uninstalled. All Registry entries and *hubiC.db* and *hubiC.db-journal* files were removed during the uninstallation process, with the exception of the application.txt file, which was located in *AppData/Local/Temp*.

The sync folder ( *C:\Users\Alice\hubiC\*), was left untouched and all files remained intact, without change to their metadata. Program installation was deleted from the *C:\Program Files* directory.

During the analysis of the memory, the e-mail address of the user was found in the memory of the current machine state**.**

## I. Analysis of Download Desktop VM (1.5.2)

The previously discovered log files and database files were analysed for notable changes that could pertain to the current activity. Additionally, the file was hashed before upload to the server, on a different machine, and hashed after download, to compare content. Analysis of the metadata, of the test file, revealed the date created field was the time when the file was uploaded to the server originally, but the access date matched the time of file synchronisation during the download.

Assessing the content of the hubiC.db file, after download completion, resulted in a new recored contained within the files table. This new file identifier, at ID 4, represented the test file used. This indicates that the hubiC.db file is a constant database representation of the contents of the sync folder. Comparing the file hashes, showed no change to the original file contents during the download process.

During the analysis of the logs, in the application.txt file in AppData/Roaming/hubiC, there were entries reporting on directories created as shown in Figure 8**.**

Figure 8 – Logs indicating folder creation

During analysis of the memory of the download VM, the downloaded file was found contained in memory, with some stray bits added.

## J. Analysis of Delete Desktop VM (1.5.2.1)

Assessing contents of the log files showed records of the deletion were available in the Application.txt file (*AppData/Roaming/hubiC*). When the test file was deleted from the sync folder, it was moved to the recycle bin, by effect of hubiC ustilising Windows Explorer to enact this deletion. Additionally, when the deletion occurs, hubiC updates the hubiC.db file located in *AppData/Roaming/hubiC*. During the memory analysis of the delete VM, the deleted file was found four times in the memory.

## IV. CONCLUSION AND FUTURE WORKS

In reference to the initial questions posted in the introduction, the following answers were gained.

The data found on the hard drive provided insight into the inner workings of the hubiC application. The Chrome Cache Viewer enabled file remnants to be obtained during the use of the hubiC web client. Username and password of the active user were found in the hubiC process memory. Additionally, the test file, deleted from the hubiC cloud service, was found, after deletion. An analysis of the network traffic provided insight into the communication occurring between the hubiC client and server, giving IP addresses of the hubiC servers and the security system used, TLSV1. These present significant risks to the secure use of the hubiC system both while the client is installed and after it has been uninstalled.

Direct future applications of this research could apply simular methodlogy to the investigation of other cloud platforms, (e.g. ADrive, eCloud, etc) on Windows 8.1. This would increase the scope of the investigation and provide greater insight to an investigator, potentially revealing common flaws across several cloud platforms. Extending this common approach to applications of other catagories of forensics techniques such as cloud malware forensics [23-27], mobile device cloud applications forensics [28,29], social network platforms investigation on cloud systems [30, 31], and cyber-crime and cyberwar investigation techniques in cloud environments [32,33] would expland this to a comprehensive body of knowledge. Finally, as forensics investigation process may reveal many private data [36,37], it is important to consider privacy effects of forensics investigation in cloud environments in future works.